%% file: main.tex
\documentclass[preprint,journal]{vgtc}       

\ifpdf
  \pdfoutput=1\relax                   
  \usepackage{graphicx}                
  \DeclareGraphicsExtensions{.pdf,.png,.jpg,.jpeg} 
\else
  \usepackage{graphicx}                
  \DeclareGraphicsExtensions{.eps}     
\fi%

\graphicspath{{figures/}{pictures/}{images/}{./}} 

\usepackage{microtype}                 
\PassOptionsToPackage{warn}{textcomp}  
\usepackage{textcomp}                  
\usepackage{mathptmx}                  
\usepackage{times}                     
\usepackage{cite}                      
\usepackage{tabu}                      
\usepackage{booktabs}                  
\usepackage{enumitem}

\usepackage{xcolor}
\usepackage{amsmath}

\usepackage[caption=false, font=footnotesize]{subfig}

\ieeedoi{10.1109/TVCG.2019.2934275}

\onlineid{1191}

\vgtccategory{Research}
\vgtcpapertype{Application/Design Study}

\newcommand{\name}{{\textit{PlanningVis}}}

\newcommand{\blue}[1]{\textcolor{black}{#1}}

\newcommand{\modified}[1]{\textcolor{black}{#1}}
\newcommand{\modifiedSecond}[1]{\textcolor{black}{#1}}

\title{\name: A Visual Analytics Approach to Production Planning in Smart Factories}

\author{Dong Sun, Renfei Huang, Yuanzhe Chen, Yong Wang, Jia Zeng, Mingxuan Yuan, Ting-Chuen Pong, and Huamin Qu}
\authorfooter{
\item
 D. Sun, R. Huang, Y. Wang, T. Pong, and H. Qu are with the Hong Kong University of Sccience and Technology. Y. Wang is the corresponding author. E-mail: \{dsunae, rhuangan, ywangct, tcpong, huamin\}@ust.hk
\item
 Y. Chen, J. Zeng, and M. Yuan are with Noah's Ark Lab, Huawei Technologies Co. Ltd. E-mail: \{chenyuanzhe, zeng.jia, yuan.mingxuan\}@huawei.com.
}

\shortauthortitle{Biv \MakeLowercase{\textit{et al.}}: Global Illumination for Fun and Profit}

\abstract{
Production planning in the manufacturing industry is crucial for fully utilizing factory resources (e.g., machines, raw materials and workers) and reducing costs. 
With the advent of industry 4.0, plenty of data recording the status of factory resources have been collected and further involved in production planning, which brings an unprecedented opportunity to understand, evaluate and adjust complex production plans through a data-driven approach. 
However, developing a systematic analytics approach for production planning is challenging due to the large volume of production data, the complex dependency between products, and unexpected changes in the market and the plant. 
Previous studies only provide summarized results and fail to show details for comparative analysis of production plans. Besides, the rapid adjustment to the plan in the case of an unanticipated incident is also not supported. 
In this paper, we propose {\name}, a visual analytics system to support the exploration and comparison of production plans with three levels of details: a plan overview presenting the overall difference between plans, a product view visualizing various properties of individual products, and a production detail view displaying the product dependency and the daily production details in related factories.
By integrating an automatic planning algorithm with interactive visual explorations, {\name} can facilitate the efficient optimization of daily production planning as well as support a quick response to unanticipated incidents in manufacturing. 
Two case studies with real-world data and carefully designed interviews with domain experts demonstrate the effectiveness and usability of {\name}. 
} 


\CCScatlist{ 
 \CCScat{K.6.1}{Management of Computing and Information Systems}%
{Project and People Management}{Life Cycle};
 \CCScat{K.7.m}{The Computing Profession}{Miscellaneous}{Ethics}
}

\teaser{
  \centering
  \includegraphics[width=0.9\linewidth]{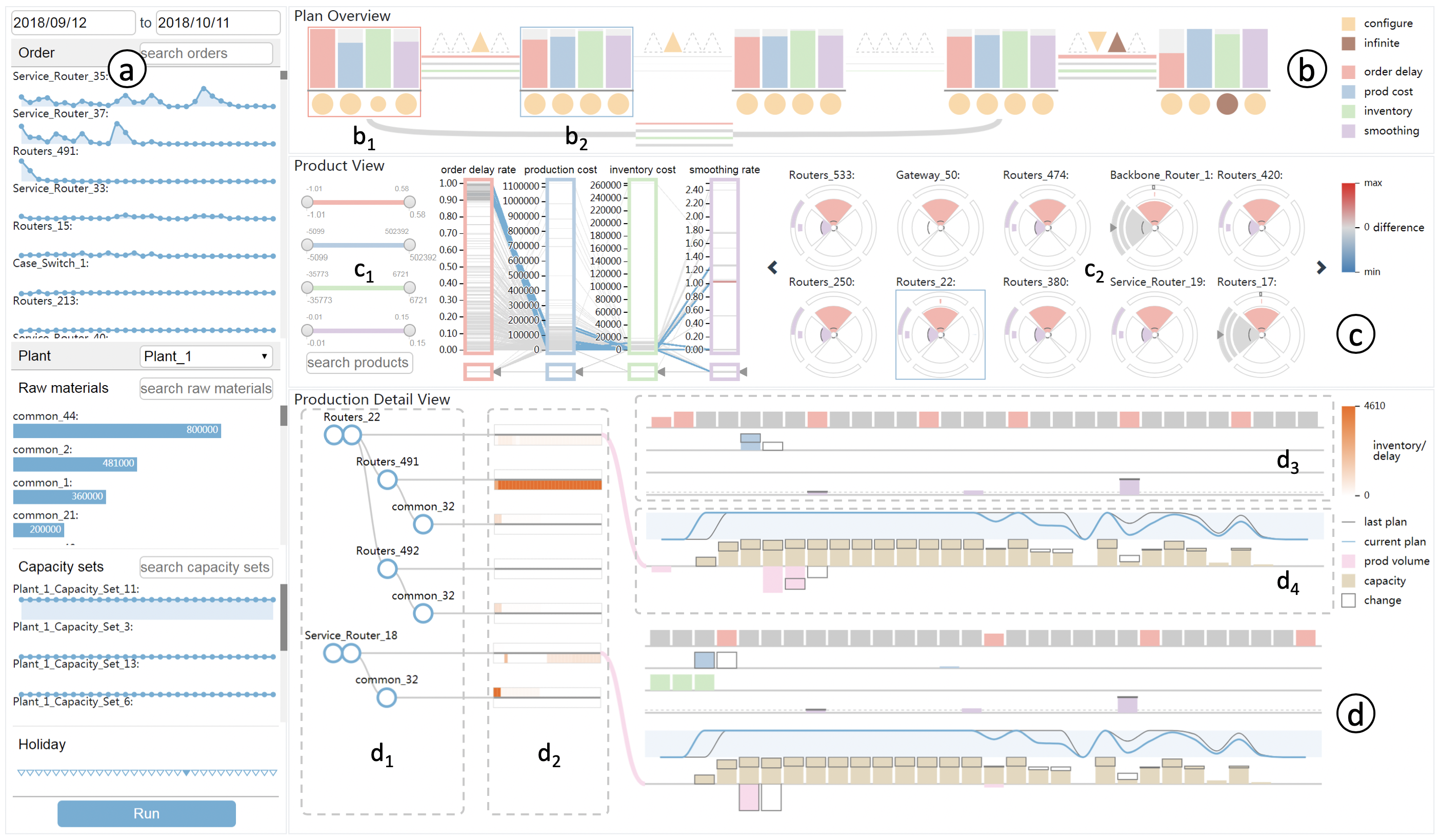}
  \caption{
  {\name} supports interactive exploration, comparison and optimization of production plans. 
  (a) The control panel enables interactively building
the configuration data of the planning algorithm.
  (b) The plan overview summarizes each plan and their differences. 
  (c) The product view reveals the distribution of all the products (c$_1$) and presents various properties of the selected products (c$_2$). 
  (d) The production detail view displays the dependency between products (d$_1$, d$_2$) and the daily production details in related plants (d$_3$, d$_4$).
  }
  \label{fig:teaser}
}

\vgtcinsertpkg

\begin{document}


\firstsection{Introduction}

\maketitle

\input{src/introduction.tex}
\input{src/background-and-requirement-analysis.tex}

\input{src/related-work.tex}

\input{src/data_modeling.tex}

\input{src/visual_design.tex}

\input{src/system_overview.tex}

\input{src/case_studies.tex}
\input{src/expert_interview.tex}
\input{src/discussion.tex}
\input{src/conclusion_and_future_work.tex}
\input{src/acknowledgements.tex}


\bibliographystyle{abbrv-doi}

\bibliography{template}
\end{document}

%% file: src/introduction.tex
Production planning in \modified{the} manufacturing industry is to assign the limited resources such as raw materials, product components, machines and workers, to different production tasks \cite{wang2001aggregate}. 
A production task specifies the type and number of products produced by each machine during a given time period, aimed at maximizing demand satisfaction and revenue. 
Therefore, production planning is critical to a manufacturing company. A good production plan should not only make full use of the limited resources to increase income, but also take the uncertainty of the market and the plant into consideration, so that it has the flexibility to be easily adjusted according to changes in the market and unanticipated problems in the plant \cite{mula2006models}. 
To ensure a cost-effective and robust production plan which takes various factors into consideration, there are three major tasks involved according to \modified{the survey by Maravelias et al.~\cite{maravelias2009integration}} and the iterative discussions with domain experts from our industry collaborator.
The first task is the resource allocation and the production assignment, which is the basis of production planning. 
The second task is to compare production plans in different levels of details, as production planning practitioners (i.e., planners) and factory managers usually need to compare plans and find the optimal one.
Finally, since changes may appear due to the uncertainty in supply, production and demand, it is also necessary to explore the adverse influence of unanticipated incidents on the production plan and decide the best adjustment strategy accordingly. 

\modified{With the advent of industry 4.0 \cite{kagermann2013recommendations} (or industrial internet \cite{evans2012industrial}), 
various sensors have been deployed on machines in the factory to record the status of equipment, and the quality of products and assembly items produced. The availability of such massive data creates new opportunities to identify production bottlenecks and make an effective production plan through data-driven approaches \cite{auschitzky2014big, posada2015visual}.}

However, it is challenging to conduct effective analysis for production planning. 
The primary challenge originates from the large amount of manufacturing data and its complexity, which makes the exploration and comparison of different production plans very difficult. For instance, in a large manufacturing company, there are usually multiple factories with various manufacturing machines and devices, tens of thousands of products and a large number of manufacturing constraints. Also, the relationships between products and machines are complex. The production of one product may rely on another product, and different products may consume the same raw materials that are processed on the same machines. 
Another challenge is to analyze the influence of a sudden change in raw material supply, the production process and the market demand. A change may affect many products and factories, which may also spread along the complex relationships among products. How to present the adverse influence with levels of details remains unknown. 
Prior work on production planning generally consists of two types: automatic algorithms and visualization approaches. 
Automatic algorithms, such as mathematical programming \cite{shapiro1993mathematical} and heuristic algorithms \cite{li2007heuristic}, have been widely utilized in production planning. These algorithms can efficiently provide a basic production plan and save many manual efforts. However, due to the complexity of manufacturing data, users may not be able to understand and adjust \modified{a} plan. Therefore, automatic approaches cannot fulfill the flexibility requirement. 
On the other hand, visualization approaches have also been employed to analyze production planning \cite{zhang1996visualizing, heilala2010developing, sydow2015visualizing, wu2001visualizing, worner2013simulation, jo2014livegantt}. These studies present the bottleneck of a production plan and support manual interactions to improve the plan. However, they only rely on manual manipulation to optimize production planning, which is inefficient and time-consuming. In addition, these approaches do not support a quick response to the sudden change in the market and the factory, which may result in a significant loss. 

In this paper, we propose {\name}, a visual analytics approach to help users combine their domain knowledge and the capability of automated algorithms. It enables fast exploration and \modified{comparison of different production plans in multiple levels of details}. Also, it supports two types of what-if analyses: the interactive optimization of production planning and a fast adjustment to the production plan in case of a sudden change in raw material supplies, the production process and the market demand.
Our approach provides three levels of details for the exploration of production planning. The \blue{plan overview (Fig. \ref{fig:teaser}b)} records the optimization history and presents the summarized algorithm results. The \blue{product view (Fig. \ref{fig:teaser}c)} shows the performance distribution of all the products, \modified{where a novel glyph design is proposed to visualize various performance indicators of individual products and the difference between two production plans.}
 The \blue{production detail view (Fig. \ref{fig:teaser}d)} demonstrates the dependency of one product on resources and other products, and the daily production in relevant plants. Rich interactions are also enabled in {\name} to help users conduct detailed exploration.
We evaluate the effectiveness of {\name} with real-world production planning data from a world-leading manufacturing company. Two case studies, including the optimization of production planning and the quick response to unanticipated incidents, and carefully designed interviews with domain experts demonstrate the effectiveness and usability of {\name}. 


%% file: src/background-and-requirement-analysis.tex
\section{\modified{Background}}
\modifiedSecond{In this section, we introduce the background of production planning, including a detailed description of the data (e.g., the performance indicators) and the analysis of tasks and requirements of production planning. The data description and requirement analysis are based on both an extensive survey of prior research and our interviews with the production planning experts from our industry collaborator.
}

\subsection{\modifiedSecond{Data Description}}
In this paper, we study the production planning problem in \modified{the} manufacturing industry. We work closely with planners and factory managers from a world-leading manufacturing company. The company owns about 50 factories performing production tasks of tens of thousands of \textbf{products} and \textbf{assembly items}.
\modified{Planners need to make a 30-day production plan every day based on a hybrid production planning algorithm \cite{sahling2009solving},
which takes the \textbf{initial inventory} of raw materials, the \textbf{production capacity} of factories~\cite{florian1971deterministic}, the \textbf{arrangement of holidays}, and the \textbf{demand of products} as the input.} 
Their work is to assign daily tasks to each factory. 
A typical production task is that Factory A is required to produce $n$ pieces of Product B on Day C. 

Each product may be produced by several factories, which jointly serve the demand. There are two kinds of demand: the \textbf{real order} from the customer, and the \textbf{predicted demand} based on past orders. In this work, we sum the two types of demand together for simplicity. 
Both the inventory and the production output of the plant can be used to serve the demand. If the demand is not satisfied, it will be delayed and produced later according to a predefined priority list of products.
 
\modified{The production in a factory relies on two factors: the production capacity of the factory and the supply of the \textbf{child components}.} 
First, the production capacity can be described as the total resources that can be used to produce products. In the same factory, different products may consume different types of production capacity, which are represented as diverse \textbf{capacity sets}. 
Second, the hierarchical structure in the dependency between products and their child components creates a \textbf{bill of materials (BOM) tree} \cite{BOM_explanation}. For example, the production of a mobile phone will consume several central processing units (CPU), a display screen, and so on, while the production of the CPU relies on the supply of arithmetic logic units (ALU), registers, and so on. The leaf nodes of the BOM tree are \textbf{raw materials}, which are purchased from other suppliers. Both the intermediate components and the raw materials on the leaf nodes of the BOM tree can be ordered by customers. 

Among multiple performance metrics in production planning, we identify four important \textbf{performance indicators} \cite{parmenter2015key} after discussions with planning practitioners and factory managers: 

\textbf{Order delay rate.}
The first and most significant objective is to minimize the order delay rate. A low order delay rate can not only increase \modified{revenue} but also improve customers' satisfaction, which can promote \modified{future sales}. 

\textbf{Production cost.}
Reducing the production cost is the second objective. A proper production can satisfy the demand for most products and eliminate inventory. 
However, a high production cost does not necessarily indicate that the production plan is bad, since it may be caused by high demand. 

\textbf{Inventory cost.}
The third objective is to minimize the inventory cost. A high inventory cost may be caused by excessive production, which will also increase the production cost. 

\textbf{Smoothing rate of production capacity use.}
As the last objective, smoothing the weekly use of production capacity aims at keeping machines in a constant working intensity. The smoothing rate is computed to represent the difference in production capacity use between two consecutive weeks. Keeping a low smoothing rate is important since the domain experts stated that the dramatically changing working intensity would cause damage to the production machine and increase the maintenance cost. 


\subsection{\modified{Task and Requirement Analysis}}
During the past six months, we have worked closely with six production planning experts from our industry collaborator. Four of them work on the production planning algorithms, while two of them are planners who need to check the algorithm outputs and further assign feasible production tasks to each factory. 

\modified{At the early stage of the collaboration, we held weekly meetings with the experts. During the meetings, they explained to us the data, the production planning algorithm, and the problems they were faced with in daily planning. 
We then applied cause and effect analysis \cite{ishikawa1990introduction} to the problems in production planning. After compiling a list of problems, we explored the potential causes based on the experts' experience and the literature review, which revealed the key information that the experts needed for decision making in production planning.}

\modified{We identified two categories of analytical tasks after detailed discussions with the experts. 
The first type of tasks focuses on the exploration and optimization of production plans. For example, \textit{What's the difference between two plans? Which products perform poorly in terms of the performance indicators? What are the key factors that restrict production? 
What kind of change in the configuration data can improve the production plan?}
The second type of tasks is related to the fast response to unanticipated changes in the market and the plant. For instance, \textit{What incidents may have an adverse influence on the production plan? What is the influence of such incidents? What strategies can reduce the adverse influence?}}

\modified{When working on this paper,} we first developed an early prototype and then improved the designs iteratively according to the feedback from the experts. At each iteration, we held weekly meetings to introduce the visual designs and collected comments from them. Finally, we formulated eight design requirements which can be grouped into four levels, and developed the current version of \modified{the} visualization system \blue{(Fig. \ref{fig:teaser})}. 

 

The \textbf{overall-plan-level} requirement focuses on providing an overview of all the production plans. 

\begin{enumerate}[label=\textbf{R{\arabic*}},nolistsep]
    \item
    \textbf{Visualize the optimization process of production planning.}
    The visual design should present the summarized algorithm results of production plans and the difference between two plans. The recorded optimization history can not only help users verify the effect of their manipulation but also provide an overview for planners to choose the best strategy. 
\end{enumerate}

The \textbf{product-level} requirements focus on displaying the statistics of individual products. 

\begin{enumerate}[label=\textbf{R{\arabic*}},resume,nolistsep]
    \item
    \textbf{Show the distribution of all the products.}
    Presenting the distribution of performance indicators for different products can reveal clusters and anomalies of products, which provides guidance for further exploration. 
    
    \item
    \textbf{Support filtering and selecting products of interest.}
    The visualization system should support interactions to filter and select products with specific performance indicator values and present detailed information about the selected products. 
\end{enumerate}

The \textbf{detail-production-level} requirements relate to \modified{the} detailed description of the production and the relationship between supply and demand. 

\begin{enumerate}[label=\textbf{R{\arabic*}},resume,nolistsep]
    \item
    \textbf{Visualize the dependency among products.}
    Due to the dependency among products in the BOM tree, the production of a parent product may be limited by the lack of child components. Showing the dependency relationship among products, along with their temporal supply/demand distribution allows problem diagnosis and production planning optimization. 
    
    \item
    \textbf{Present the production detail of a product in different factories.}
    The system should display the daily production output and production capacity use of each factory. The analysis of detailed production can disclose the workload of each plant and the reason for insufficient production. 
\end{enumerate}

The \textbf{comparison-level} requirements aim at comparative analysis and optimization of production plans. 

\begin{enumerate}[label=\textbf{R{\arabic*}},resume,nolistsep]
    \item
    \textbf{Enable interactive optimization of the production plan.}
    The domain experts are eager for the support of visual interactions to improve production planning. To this end, our design should combine the automated algorithm and domain knowledge. The visual encoding should provide guidance for the manipulation and rapid feedback is needed to verify the effect. 
    
    \item
    \textbf{Support a fast response to unanticipated incidents.}
    A sudden change in the market and the plant may have an adverse influence on production planning. Revealing the influence and supporting a quick adjustment are critical to developing an efficient production plan. 
    
    \item
    \textbf{Support comparative analysis of two production planning strategies.}
    The comparison of production plans is important in the checking of the effect of optimization and showing the impact of unanticipated changes in the market and the plant. The visualization system should provide \modified{a} detail-on-demand comparison of planning strategies for decision making. 
\end{enumerate}

%% file: src/related-work.tex
\section{Related Work}
The related work can be categorized into two types: manufacturing data visualization and time-series data visualization. 


\subsection{Manufacturing Data Visualization}
Many visual analytics approaches have been proposed for exploring manufacturing data. According to the survey by Ramanujan et al. \cite{ramanujan2017visual}, there are mainly two types of work: visualization for production planning and simulation, and visualization for process monitoring. 

\textbf{Visualization for production planning and simulation.}
Previous work on production planning and simulation aims at revealing the bottleneck, identifying potential problems, and supporting interactions to improve the production strategy. 
VIZ\_planner \cite{zhang1996visualizing} is an early study which utilizes bar charts to visualize multiple attributes of production planning. 
Sydow et al. \cite{sydow2015visualizing} used Sankey diagrams \cite{wongsuphasawat2012exploring} to show the relationships between orders and available resources. 
\modified{These studies provide useful insights into production planning, but they focus on the production planning of a small number of products and are not applicable to the production planning of a rather large number of products.}
Wu et al. \cite{wu2001visualizing} developed a visualization system to reveal weekly machine \modified{loads} in the metal ingot casting process. It supports manually moving the production task from one week to another.
LiveGantt \cite{jo2014livegantt} extends the Gantt chart with reordering and aggregation algorithms to identify common scheduling sequences and support \modified{the} rearrangement of production tasks. 
\modified{However, these studies mainly rely on the manual adjustment of the production plan and fail to take advantage of automatic algorithms.}
Worner et al. \cite{worner2013simulation} provided an interactive visual interface where planners could redesign the manufacturing layout. Immediate feedback is provided to support the comparison of different strategies. 
\modified{However, the work only provides overall performance indicators for the comparison of manufacturing layouts.}
The simulation of production processes has also been studied to identify anomalies and planning drawbacks for decision making. 
Zhou et al. \cite{zhou2011visualizing} applied visualization techniques to the simulation of steel manufacturing to promote the understanding of the complex manufacturing system. W{\"o}rner et al. \cite{worner2011visual} visualized the simulation run of the assembly line to identify potential congestion during the manufacturing process. Post et al. \cite{post2017user} developed a visualization system to show temporal changes in the simulated production process. 


\textbf{Visualization for process monitoring.}
With the advent of industry 4.0, a great number of manufacturing data have been collected by sensors deployed in the plant, which promotes the adoption of visual analytics approaches to process monitoring. 
TTPView \cite{matkovic2002process}, as an early study, employs focus+context dashboard visualization techniques to present the process monitoring data. ViDX \cite{xu2017vidx} adopts an outlier-preserving aggregation approach in the Marey's graph \cite{tufte2001visual} to identify patterns and diagnose problems in assembly lines. It supports user-driven anomaly detection and presents the 3D model of the machine with errors. BlueCollar \cite{herr2019bluecollar} visualizes the path of the workers in a manufacturing site to facilitate the optimization of production layouts. Chen et al. \cite{chen2018sequence} developed a new sequence mining technique to summarize the patterns of vehicle fault records. Wu et al. \cite{wu2018visual} integrated advanced algorithms with visual analytics approaches to monitor the equipment and predict risks. Zhou et al. \cite{zhou2018visually} visualized the running status of manufacturing facilities by a matrix-based heatmap. 

\modified{Although previous studies on production planning can help planners recognize patterns, discover bottlenecks, and improve planning strategies, 
they only offer several overall performance indicators for the comparison of different production plans. Planners have no idea about the attributes of each product and the production details in each plant. 
Besides, the influence of a sudden change in the raw material supply, the production process and the market demand is not considered, which may reduce the flexibility of a production planning strategy and result in a significant loss.}
Compared to previous work, our approach enables detail-on-demand exploration and comparison of different planning strategies. It combines domain knowledge and automated algorithms to support efficient optimization of production planning.  
Furthermore, our solution can reveal the adverse influence of unanticipated changes in the market or the plant, and facilitate a quick adjustment to the production plan.  

\subsection{Time-series Data Visualization}
A number of techniques have been developed for time-series data visualization. The existing work can generally be classified into two categories based on the arrangement of the time domain \cite{aigner2011visualization, brehmer2017timelines}: linear time-series data visualization and cyclic time-series data visualization. 

\textbf{Linear time-series data visualization.}
Most of the time-series data have a linear time domain, where each time primitive has different predecessors and successors. 
An early and widely-used approach for presenting linear time-series data is line charts \cite{playfair1801commercial}. Based on line charts, small multiples \cite{tufte2001visual} were proposed for overview and comparison. Horizon graphs \cite{saito2005two}, stacked graphs \cite{byron2008stacked}, and braided graphs \cite{javed2010graphical} display data in a compact form and make it possible to present a large amount of data \modified{on} a limited screen. 
With the increase of the complexity and volume in time-series data, multiple visual analytics systems have been designed to perform various analysis tasks. Line Graph Explorer \cite{kincaid2006line} supports the exploration of large-scale time series data by embedding focus+context encoding into line charts. It encodes values by \modified{the color} instead of \modified{the height}, thus providing a space-saving overview. Javed et al. \cite{javed2010stack} proposed stack zooming, a multi-focus method for exploring long time-series data, which was later improved by KronoMiner \cite{zhao2011kronominer} and Timenotes \cite{walker2016timenotes}. Another multiresolution method is MultiStream \cite{cuenca2018multistream}, which extends the basic streamgraph to describe hierarchical data. 

\textbf{Cyclic time-series data visualization.}
Cyclic time domain is composed of periodic time primitive sequences. 
Calendar-based visualization \cite{van1999cluster} is used to describe data with weekly, monthly, or yearly patterns. Another visualization technique, the spiral graph \cite{weber2001visualizing}, utilizes rings to represent cyclic time-series data. 
Recently, more specialized visualization systems are developed for periodic time-series data. For example, T-Cal \cite{fu2018t} visualizes team communication data with a calendar-based approach. IDMVis \cite{zhang2019idmvis} shows the discrete records of a diabetic's physical condition with colored dots to promote the reasoning and refinement of the treatment. 

Our work is inspired by the techniques for visualizing linear time-series data. The production planning data are essentially time-series data. We integrate existing time-series data visualization techniques with the visual encoding for comparative analysis to support the exploration of production plans. 

%% file: src/data_modeling.tex
\section{Data Modeling}
This section first introduces the hybrid production planning algorithm. Then, the output of the algorithm is processed for feature extraction and comparative analysis. 

\subsection{The Hybrid Production Planning Algorithm}
The production planning problem is inherently a multi-objective optimization problem. 
\modified{
Specifically, we are focusing on minimizing the four performance indicators suggested by the domain experts: the order delay rate, the production cost, the inventory cost and the weekly smoothing rate of production capacity use.
} 

\modified{Then, we employ the weighted sum model (WSM) \cite{triantaphyllou2000multi} to define the multi-objective production planning problem. The WSM can be solved by integer programming.}
However, directly employing integer programming to solve such a large scale production planning problem is time-consuming. Therefore, in this paper, we adopt a hybrid production planning algorithm which combines linear programming \cite{shapiro1993mathematical} and heuristic algorithms \cite{li2007heuristic}. It is adapted from an approach proposed by Sahling et al \cite{sahling2009solving}.

\subsection{Data Processing}

After getting the production planning data generated by the automatic algorithm, further data processing will be conducted to extract features and analyze the difference between two plans. 

\modified{
We first compute the daily performance indicators of each product, including the order delay rate, the production cost, the inventory cost, and the smoothing rate of production capacity use. 
These performance indicators are used to evaluate the production plan and identify products with production problems.}
Also, we calculate the summarized statistics such as the mean and the variance of a product over 30 days. 

The next step is to preprocess two types of anomalous data we identified: missing data and infinite values. 
The data of some products are missing because there is no demand for these products, they do not consume any capacity set, or the products are not involved in the production process at all. We assign special negative values to these products so that they can be recognized and displayed differently in the visual design. 
There also exist some infinite values in the smoothing rate of production capacity use, owing to the addition of new capacity sets or the signing of the contract with new factories. After the discussion with domain experts, we decide to replace these infinite values with a suitable number, which is neither large enough to affect the summarized performance indicators much  nor too small to be noticed in the visualization system. 

The last step is to reveal the difference between two production plans, with the aim of supporting the optimization of production planning, and showing the impact of unanticipated changes in the market or the plant. 
We summarize three levels of differences: the plan level, the product level, and the production detail level.

%% file: src/visual_design.tex
\section{Visual Design}
The {\name} system is developed to support multi-level exploration and comparison of different production plans, which consists of four parts: a control panel \blue{(Fig. \ref{fig:teaser}a)} to present the configuration data and promote interactively changing the configuration, a plan overview \blue{(Fig. \ref{fig:teaser}b)} to exhibit the optimization history and compare \modified{the} summary information of different plans, a product view \blue{(Fig. \ref{fig:teaser}c)} to reveal the distribution of products and explore individual products of interest, and a production detail view \blue{(Fig. \ref{fig:teaser}d)} to display the BOM tree of products and the production detail of the selected product in related plants. 
It combines automatic algorithms and domain knowledge to enable two kinds of what-if analyses: the optimization of production planning, and the simulation of unanticipated changes to facilitate fast re-planning. 
The proposed visualization system targets at eight requirements which can be grouped into four levels (Section 2.2). 

In this section, we first describe each component of the {\name} system as well as the design alternatives. Then, we illustrate the user interactions provided to enable the exploration and analysis of production planning. 

\subsection{Control Panel}
The control panel \blue{(Fig. \ref{fig:teaser}a)} aims at supporting visual manipulation on the configuration data of production planning to generate new plans. It displays two types of configuration data: the daily demand of each product, and the available resources in each factory, including the initial inventory of raw materials, the capacity sets and the holiday arrangement. 
Based on the interactions provided by the control panel, users can leverage their knowledge to improve a production plan (\textbf{R6}) and simulate an unanticipated incident so that they can take measures in time to reduce the influence (\textbf{R7}). 

\textbf{The visual encoding of the configuration data.}
The initial order demand and the resource configuration of the selected plant will be shown when users specify the start and end dates of production planning. 
The order demand for a product is displayed by a line chart overlaid with an area chart, where the horizontal axis represents the time and the vertical axis encodes the value. Additionally, the small circles on the line chart can be dragged to change the value. 
For the production resources in a plant, we encode the initial inventory of the raw material by a draggable bar chart and illustrate the capacity sets in a similar manner to the order demand for products. 
Furthermore, the holiday schedule is represented by small triangles, where blue filled triangles indicate holidays. Users can click an unfilled triangle to arrange a holiday or click a blue filled one to cancel the holiday. 

After modifying the configuration data, users can click the ``Run'' button to invoke the production planning algorithm, and the returned result will be added to the plan overview. 

\subsection{Plan Overview}
The plan overview utilizes timeline-based glyphs to present the summarized information of various production plans and their differences. The view can reveal the macroscopic impact of configuration changes, including improving the plan and simulating unanticipated incidents in the market or the plant (\textbf{R6}, \textbf{R7}). In addition, it displays the recorded planning history which enables users to progressively optimize the plan (\textbf{R1}). The visual design is composed of plan glyphs and the links between them. 
For visual comparison \cite{gleicher2011visual}, we adopt juxtaposition between plan glyphs and explicit encoding in links (\textbf{R8}). 

\begin{figure}[tb]
    \centering
    \subfloat[Plan glyph\label{3a}]{%
       \includegraphics[width=0.32\columnwidth]{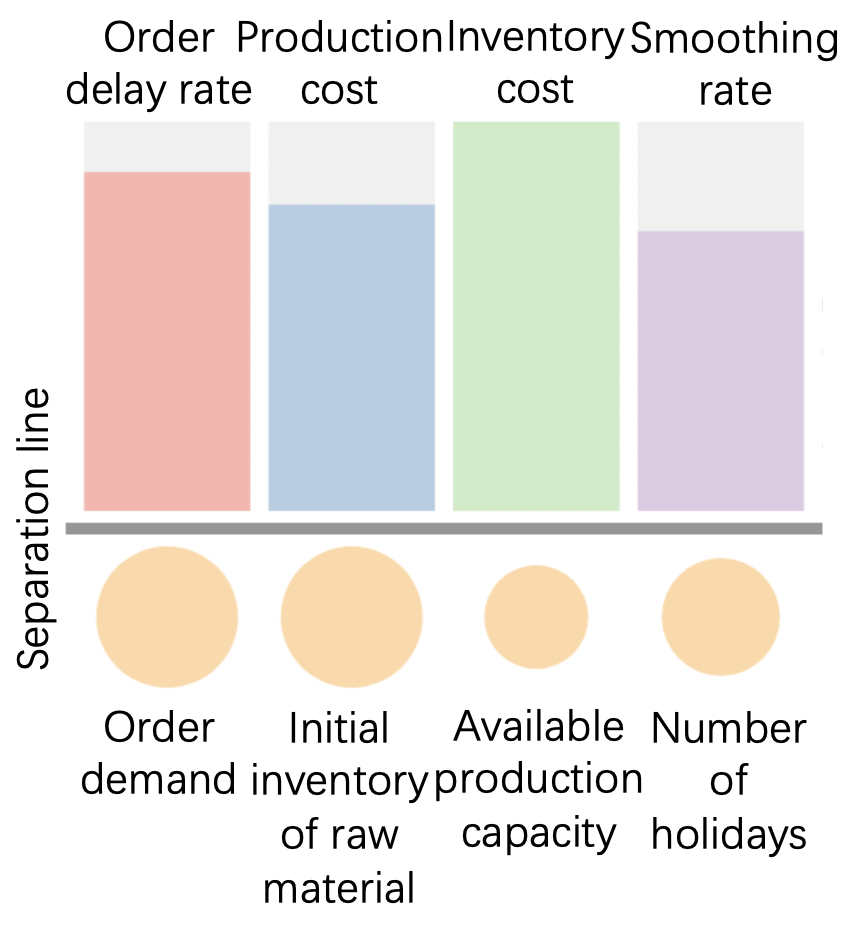}}
    \hfill
    \subfloat[Star plot\label{3b}]{%
        \includegraphics[width=0.32\columnwidth]{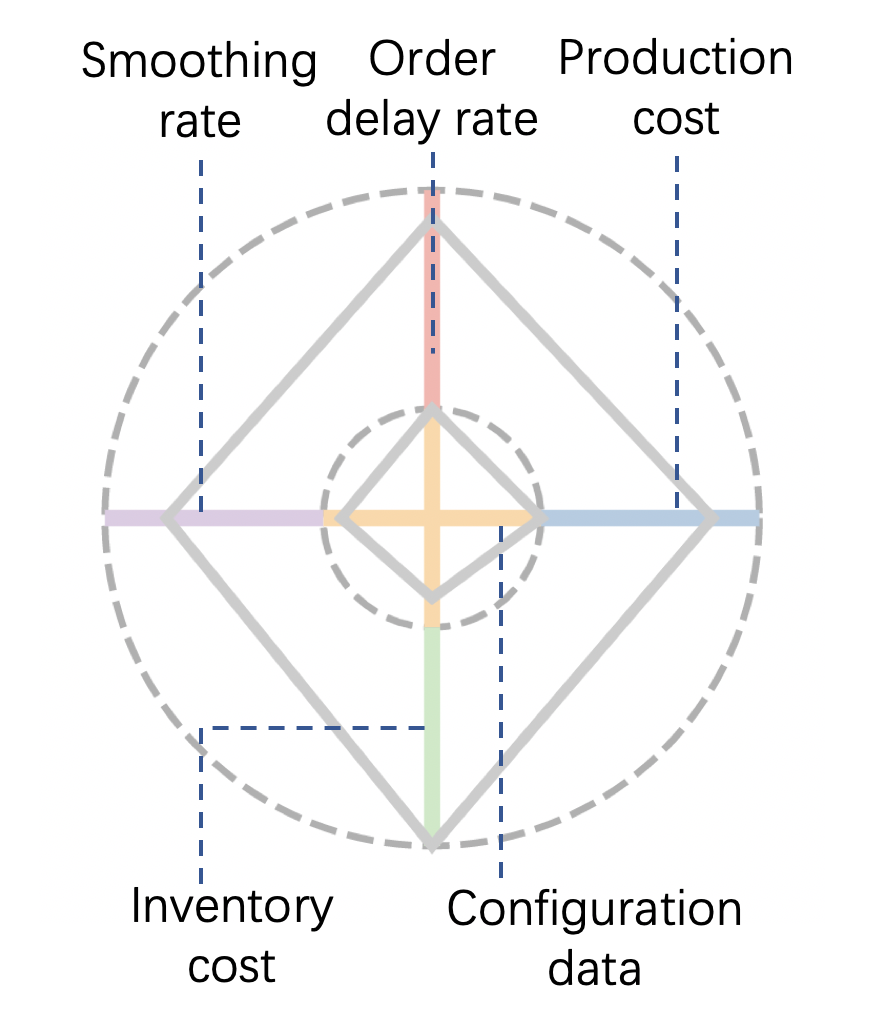}}
    \hfill
    \subfloat[Treemap\label{3c}]{%
        \includegraphics[width=0.32\columnwidth]{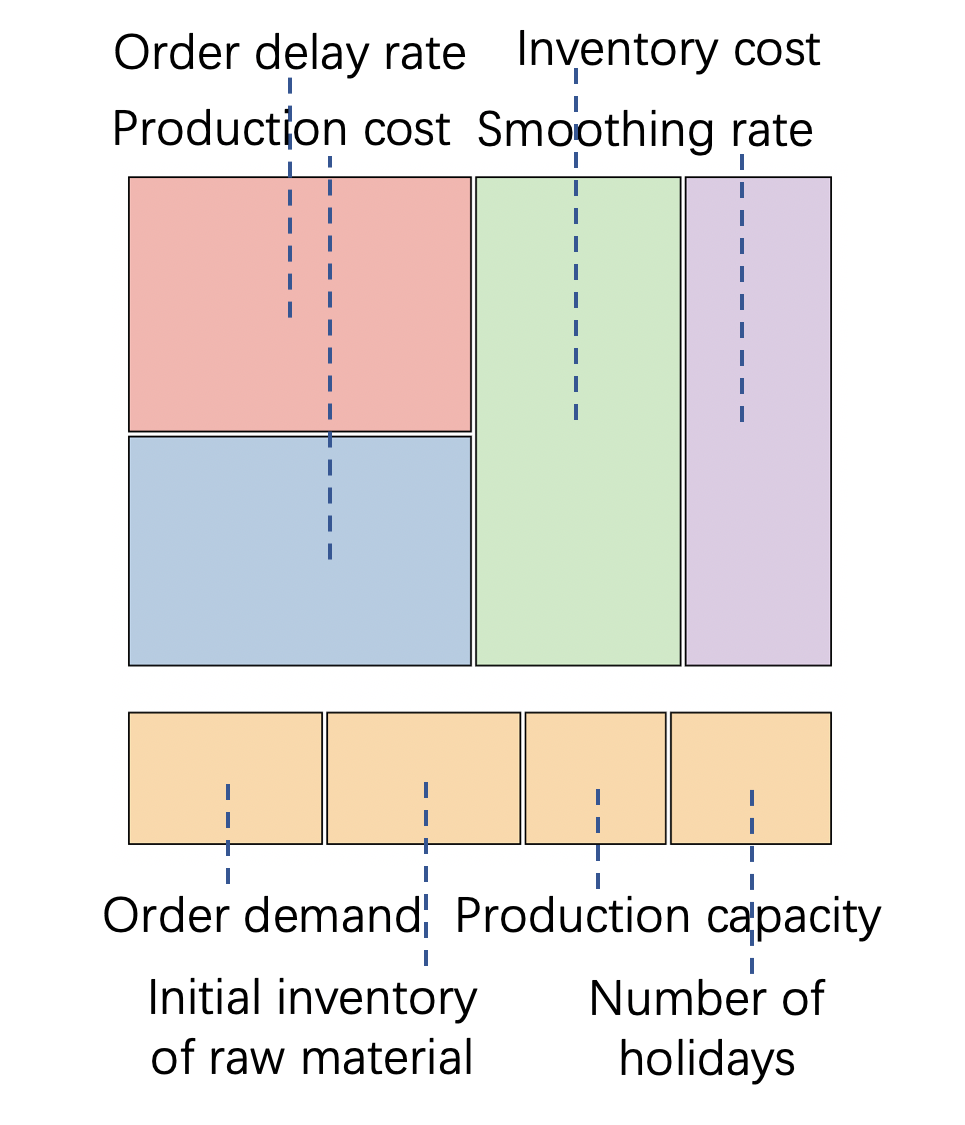}}
    \caption{
    The glyph design to represent a plan in the plan overview. (a) The glyph design employed in {\name}. (b) A star-plot based glyph design. (c) A treemap-based glyph design. 
    }
    \label{fig:plan_glyph}
\end{figure}

\textbf{Plan glyph.}
The plan glyph encodes summarized algorithm configuration data and performance indicators, as illustrated in \blue{Fig. \ref{3a}}. 
The upper part is a bar chart that shows the four key performance indicators suggested by the domain experts, including order delay rate (red), the production cost (blue), the inventory cost (green), and the smoothing rate of production capacity use (purple), where the color scheme is consistent with other views. The \modified{light} gray background in the bar chart indicates the maximum value among all the plans. 
\modified{The lower part uses light orange circles to represent four kinds of configuration data.
}
These circles, from left to right, indicate the order demand, the initial inventory of raw materials, the available production capacity, and the number of holidays, respectively. The radii of these circles describe values in disparate ranges. 
In practice, users may assume the production capacity is infinite so that they can focus on the analysis of other production constraints. \modified{For this special case, we visualize the infinite production capacity as a brown circle (the last plan glyph in \blue{Fig. \ref{fig:teaser}b})}.
The upper part and the lower part are separated by a horizontal line to avoid the misunderstanding that a relationship between the vertically aligned visual primitives exists. 

\textbf{Plan glyph design alternatives.}
Two alternatives of the plan glyph are considered during the iterative design process: the star plot and the treemap. 
As shown in \blue{Fig. \ref{3b}}, the star plot employs four spokes to display different variables, where the inner orange part encodes the configuration data, while the outer part encodes the performance indicators.
However, a deviation in interpreting the shape of the star may occur, which has been studied in previous work \cite{klippel2009star}, thus making it difficult to compare plans with small differences. In addition, users may mistakenly think that the configuration variable and the performance indicator in the same spoke have a close relationship. 
The treemap design is depicted in \blue{Fig. \ref{3c}}, where the bottom orange part exhibits the normalized configuration data and the top part exhibits the normalized performance indicators. 
However, it is not convenient to compare different variables in two production plans in the treemap. 

\textbf{The visual encoding of the links between plan glyphs.}
The links, as illustrated in \blue{Fig. \ref{fig:teaser}b}, connect two plans in the optimization history to display the difference between them.
\modified{The triangles on the upper part of the link represent the changes of the four types of configuration data, which follow the same order as that of the plan glyph.}
The size of the triangle encodes the value while the orientation of the triangle describes the increase (up) or the decrease (down) of the value. A triangle with a dashed border means that there is no change \modified{in} this type of configuration data. 
We use four horizontal lines on the lower part to show the change in the performance indicators. The width of the lines \modified{encodes} the value and the order is the same as that in the plan glyph. The same color scheme is used to indicate \modified{a} decrease while a gray line means \modified{an} increase. 

We enable users to hover over the visual cues to view the detailed information and delete a plan. Users can also click the plan glyph to choose the last plan and the current plan for exploration and comparison. An additional link will be shown at the bottom when the selected plans are not consecutive (\blue{Fig. \ref{fig:teaser}b}). After selecting the plans, the product view will give a detailed description of the products.

\subsection{Product View}
The product view contains two components: a segmented parallel coordinates plot to reveal potential clusters and anomalies of products, and the product glyph to give a detailed description of filtered products. 
The view can display the distribution of all the products (\textbf{R2}) and support filtering and selecting for further exploration (\textbf{R3}). It provides mesoscopic information on the product level for comparative analysis of two plans (\textbf{R8}), and gives support to the improvement and simulation of production planning (\textbf{R6}, \textbf{R7}). 

\textbf{The segmented parallel coordinates plot.}
Since we use a special negative value to represent the product with no demand (Section 4.2), traditional parallel coordinates plots will create a large gap between the normal values and the abnormal ones, and thus compress normal values into a small area. To resolve this issue, we design a segmented parallel coordinates plot. 
As illustrated in \blue{Fig. \ref{fig:teaser}c$_1$}, we extend the axis in \modified{the} traditional parallel coordinates plot to two rectangles. The four pairs of rectangles represent the four performance indicators, respectively. Within each pair of rectangles, the upper part displays normal values while the lower part displays abnormal values. Each line in the plot represents a product. When the line passes through a rectangle, a red-and-blue color encoding of this line segment refers to the difference between the last plan and the current plan. The red color indicates an increase and blue indicates a decrease. A gray triangle on the right side of the bar shows that the performance indicator of the current product has an abnormal value in the last plan. Additionally, a brush on the rectangle will highlight the selected products and their product glyphs will be displayed for further exploration.

On the left of the parallel coordinates plot, we also provide sliders which show the ranges of differences for users to filter products. The user can also search for products of interest. 

\textbf{Design alternatives.}
Two candidate designs are discussed to display the distribution and clustering of products. 
The first design is a scatter plot with the layout generated by the multidimensional scaling (MDS) \cite{kruskal1964multidimensional} technique. However, the domain exports report that they find it difficult to understand why certain products form a cluster, and they cannot identify the distribution of products on the four performance indicators. 
Another alternative design is the traditional parallel coordinates plot. We extend it to handle abnormal products and display the difference between plans as discussed above.

\begin{figure}[tb]
    \centering
    \includegraphics[width=\columnwidth]{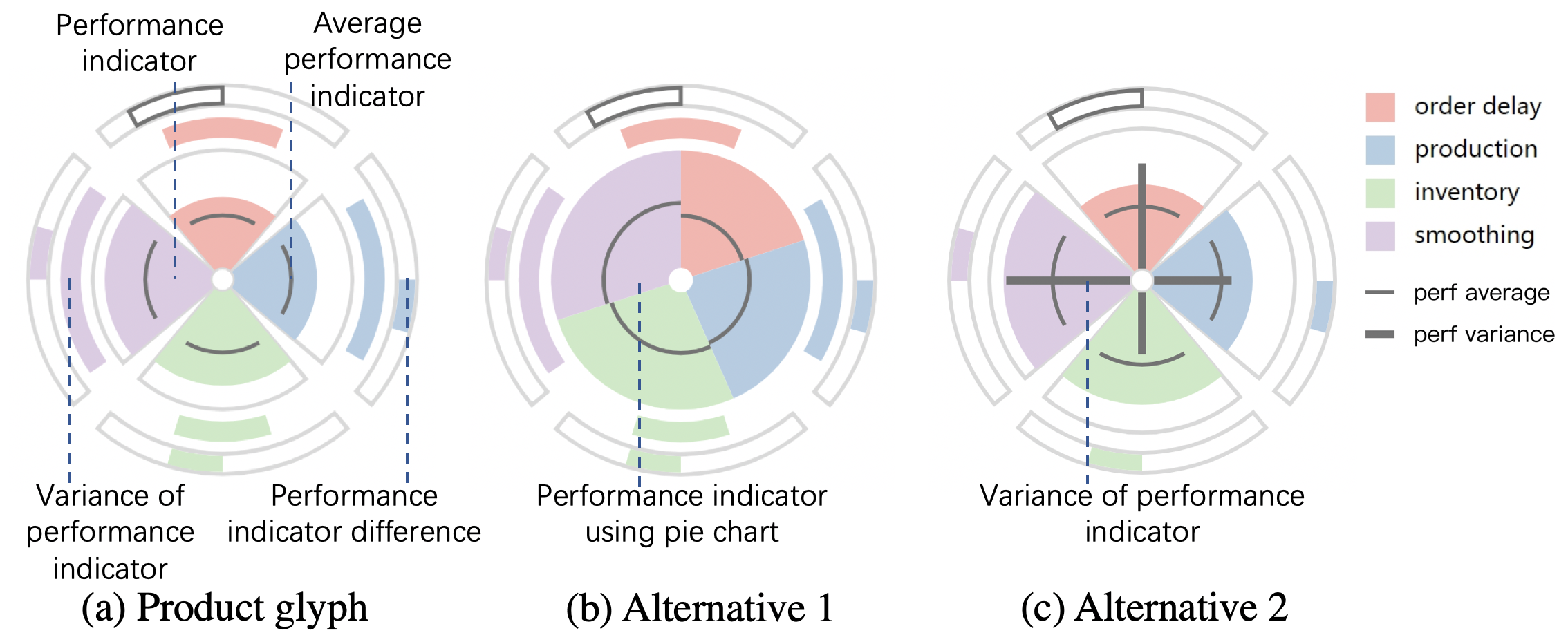}
    \caption{
    (a) The glyph design to describe various properties of a product. (b, c) Two alternative solutions to the product glyph. 
    }
    \label{fig:product_glyph}
\end{figure}

\textbf{Product glyph.}
In \blue{Fig. \ref{fig:product_glyph}a}, the product glyph depicts the selected products in detail. It has a circular shape and is tangentially partitioned into four regions to present different performance indicators. 
The radius of the innermost sector encodes the value of the performance indicators, and the black line on it points out the average value of all products. 
The angle of the arc in the middle shows the variance of the daily performance indicator during the 30-day production planning period. 
We also present the differences in the performance indicators between the last plan and the current plan in the outer arc. The arc starts from the center of each region, and the angle shows the difference. An arc \modified{extending} clockwise with the same color scheme to the performance indicator expresses an increase, while one \modified{extending} counterclockwise with only a black and bold border expresses a decrease. 
A special case is the data with predefined negative values (Section 4.2), which are shown in gray in all the visual elements. Additionally, we use a gray triangle at the outermost part of the glyph to indicate that the corresponding performance indicator of the last plan is a special negative number. 
The user can hover over the visual cues to see the numerical values and click on the product glyph to explore the production details of this product. 

\textbf{Product glyph design alternatives.}
\modified{Three other alternative designs are also considered before we finalize the current glyph design.
} 
\blue{Fig. \ref{fig:product_glyph}b} shows one choice which utilizes \modified{a} traditional pie chart to compare different performance indicators of one product. However, it is not easy to compare the same performance indicator between products because the corresponding sectors may have diverse start angles and end angles.
Another choice is shown in \blue{Fig. \ref{fig:product_glyph}c}. It uses the length of the line along the radius of the sector to encode the variance. However, since the line and the sector have different scales, overlaying them may confuse the user. 
\modified{The box plot is also considered here. Although a box plot can show the distribution of one performance indicator over 30 days, it is not space efficient to show four box plots to represent the performance indicators of one product. }

\subsection{Production Detail View}
The production detail view can be divided into the left and right parts. The left side is a dependency tree visualization, which reveals the dependency between products (\textbf{R4}). The right side \modified{is} extended bar charts, which visualize the daily production information of the selected product in related factories (\textbf{R5}).
The visual design discloses the microscopic differences between two production planning strategies (\textbf{R8}). It can further help guide the improvement of the production plan (\textbf{R6}) and reveal the impact of an unanticipated incident (\textbf{R7}). 

\textbf{Dependency tree.}
As illustrated in \blue{Fig. \ref{fig:teaser}d$_1$}, the dependency tree describes production dependency from the parent of the selected product to raw materials. The tree layout is generated by a depth-first search (DFS) algorithm \cite{tarjan1972depth} starting \modified{from} the node of the selected product, where the vertical position of the product is ranked by the access order and the horizontal position is arranged by the depth of the node in the tree. The parent node of the selected product is placed at the top of the tree. The node with children can be folded and unfolded upon clicking, thus helping the user explore a large BOM tree. 
Each unfolded node is accompanied by a heatmap vertically aligned at the right side of the node (\blue{Fig. \ref{fig:teaser}d$_2$}). The heatmap contains two rows to reveal the relationship between supply and demand. The upper row of the heatmap encodes the daily remaining inventory, while the lower row encodes the daily delayed order. The value is represented as the color saturation in the two rows. All the heatmaps are horizontally aligned. In this way, by viewing the tree with heatmaps, users can identify the impact caused by the shortage of the child product to the parent one. 
When a user clicks on the heatmap, the daily performance indicators and the production in related plants of the clicked product will be presented. 

\textbf{The visual encoding of daily production information.}
\blue{Fig. \ref{fig:teaser}d$_3$} displays the daily performance indicators, namely, the order delay rate, the production cost, the inventory cost, and the smoothing rate of production capacity use, from top to bottom. We encode the value of the current plan with the height of the bar which follows the same color scheme used before. The black rectangular border is used to encode the difference between the current plan and the last plan. The value of the current plan is larger when the border is within the colored bar, and smaller when the border is outside the bar. Bars will be changed to gray if the value is a special negative number we set before, which means the raw data is missing (Section 4.2). Note that maybe only a part of \modified{the} bars representing the order delay rate is gray since the demand is zero on these days. There are only four bars for the smoothing rate of production capacity use because it is computed once a week. A dashed line is shown to point out where the value is zero. 
In \blue{Fig. \ref{fig:teaser}d$_4$}, a line chart and a bar chart are utilized to show the daily production in a plant. The clicked product in the dependency tree and the related plants are connected by curves, whose width and color saturation represent the total production output in that plant. 
In each plant, the downward bars indicate the daily production output of the product and the upward bars indicate the use of the corresponding capacity set. The visual design is similar to that of the performance indicators. Here, the upward bars may be changed to gray, which implies that the production capacity for this product is infinite. 
Above the bar chart, a black line encodes the capacity utilization rate of the last plan and a blue one encodes that of the current plan. 


\subsection{User Interactions}
{\name} provides various interaction methods to support the exploration and comparison of production plans, the daily optimization of planning, and the quick response to unanticipated changes in raw material supply, the production process and market demand.

\textbf{Navigating through different levels of details.}
We provide three levels of details for users to explore production planning data: the plan level, the product level, and the production level. 
The user can first get the summarized information of each plan from the plan overview. 
He can then select any pair of plans to browse the product-level difference and explore various properties of individual products in the product view. The detailed production dependency and the temporal statistics are displayed in the production detail view when he clicks one product.  

\textbf{Modifying configuration data.}
In the control panel, interactive manipulation is supported to change the value of the configuration data. For order demand and capacity sets, the value can be modified by dragging the dot on the line. For raw materials, users can drag the bar to specify a new value. Besides, a click on the triangle can change the arrangement of the holiday. 

\textbf{Filtering, brushing, linking and highlighting.}
In the product view, users can filter the change of performance indicators between two plans to focus on products of interest. After that, the products selected by brushing on the segmented parallel coordinates plot will be highlighted, which will also be displayed in individual glyphs for further exploration. Meanwhile, the capacity sets in the control panel will also be filtered accordingly.
In addition, we provide search boxes in the control panel and the product view to help users look up specific orders, raw materials, capacity sets, and products. 
An informative tooltip will also be displayed when users hover over a visual element in {\name}. 


%% file: src/system_overview.tex
\section{\modifiedSecond{System Architecture and Implementation}}

\begin{figure}[tb]
 \centering
 \includegraphics[width=\columnwidth]{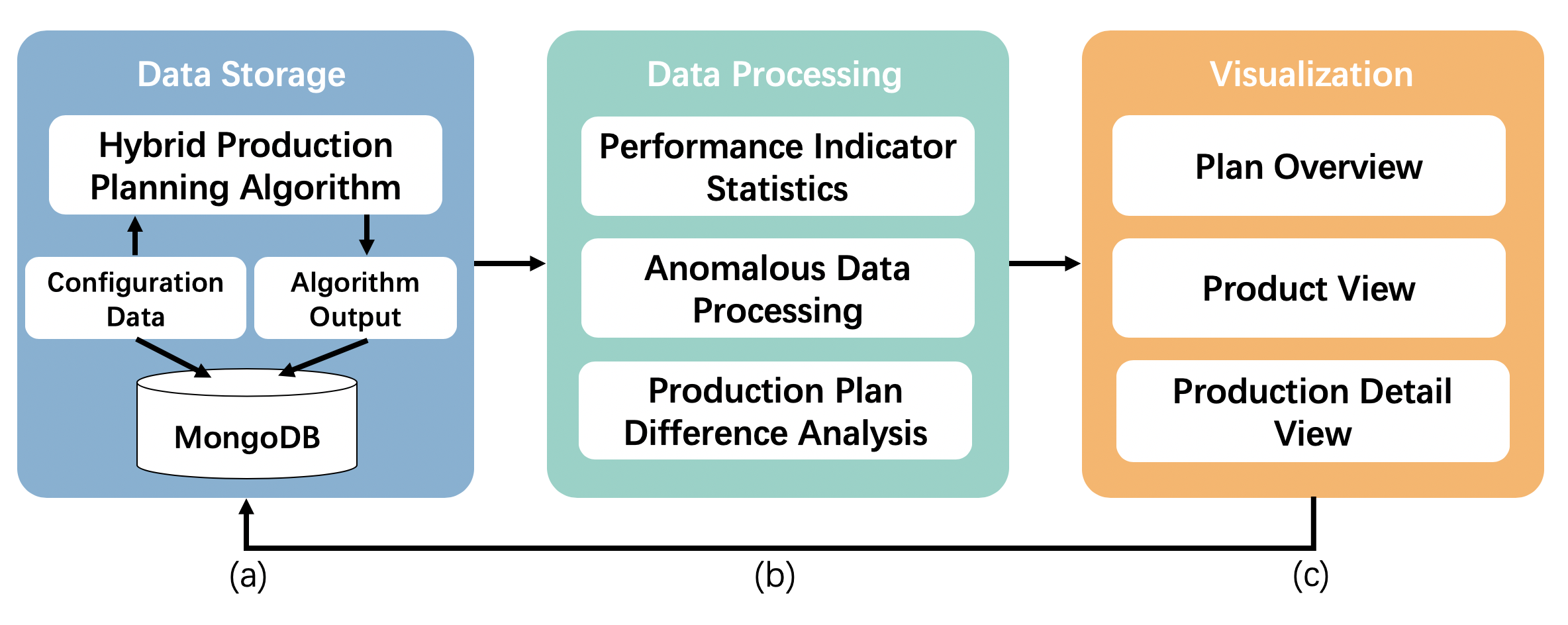}
 \caption{
 System architecture. {\name} consists of three major modules: (a) data storage, (b) data processing and (c) visualization. The data storage module collects and stores the configuration data and the result of a hybrid production planning algorithm. The data processing module pre-processes the anomalous data, computes the performance indicators and evaluates the difference between two production plans. The visualization module supports interactive exploration and comparison of production plans through three well-coordinated views.  
 }
 \label{fig:system_overview}
\end{figure}

\modifiedSecond{Fig.~\ref{fig:system_overview} illustrates the architecture of the {\name} system, which consists of three modules: data storage, data processing, and visualization.}
The data storage module collects the configuration data and the result of a hybrid production planning algorithm (Section 4.1), and \modified{stores them on a server with 64 Intel Xeon CPU processors (E7-4820, 2.0 GHz) and 256 GB memory.} 
The data processing module mainly pre-processes anomalous data, computes the performance indicators, and calculates the difference between two production plans based on user requirements. \modified{It is deployed on another server with 64 Intel Xeon CPU processors (E7-4820, 2.0 GHz) and 512 GB memory.}
The visualization module combines three well-coordinated views to support \modified{the} exploration and comparison of production plans in multiple levels of details, \modified{which is displayed on a 23.8 inch monitor with a resolution of 1920 x 1080.}
\modified{The back-end of the {\name} system is supported by Flask\footnote{http://flask.pocoo.org/}, where MongoDB\footnote{https://www.mongodb.com/} is used for data storage and Pandas\footnote{https://pandas.pydata.org/} is employed for data processing. The front-end of the {\name} system is implemented by Vue.js\footnote{https://vuejs.org/} and D3.js \cite{bostock2011d3}.}

When using {\name}, users can first configure the input data of the model based on their needs (e.g., coping with unexpected manufacturing events) in the control panel \blue{(Fig. \ref{fig:teaser}a)} and run the production planning algorithm. Then, the results of the production planning algorithm will be further visualized with three levels of details \blue{(Figs. \ref{fig:teaser}b, \ref{fig:teaser}c and \ref{fig:teaser}d)} to facilitate \modified{the} quick exploration and comparison of different production plans.



%% file: src/case_studies.tex
\section{Case Studies}
\modified{We conducted two case studies to demonstrate the effectiveness of {\name}. The users involved in our case studies are domain experts from our industry collaborator, as will be introduced in Section~\ref{sec_expert_interview}. We used a sampled dataset from the real production planning of telecommunication equipment. It contains the 30-day planning data of 1038 products or assembly items.
This sampled dataset is used in both the case studies and the subsequent expert interview (Section~\ref{sec_expert_interview}).}
%
%

 
\subsection{The Optimization of Daily Production Planning}
\modified{In the first case study, the users were asked to identify potential problems in a production plan and further optimize it (\textbf{R6}). During this exploration process, their actions, findings and comments were recorded.}
The users started with an initial production plan (\blue{Fig.~\ref{fig:teaser}b$_1$}) and explored different levels of details based on the visualization system (\textbf{R1}). When they brushed the products with high order delay rates in the parallel coordinates plot, the glyphs of the selected products were displayed on the right and the control panel listed the capacity sets related to the production of the selected products (\textbf{R3}). 
The users then browsed through the product glyphs and clicked the glyphs to view the BOM tree and daily production in related factories (\textbf{R5}). They found the line charts were close to the top, which indicated the maximum production capacity, meaning that production capacity use was saturated. This visual cue hints that the production bottleneck is due to the lack of production capacity. They then hovered over the plant and found the saturated production capacity was mainly \textit{Plant\_1\_Capacity\_Set\_11} and \textit{Plant\_1\_Capacity\_Set\_3}. Therefore, the users tried to improve the plan by increasing these capacity sets.
\begin{figure}[tb]
    \centering
    \includegraphics[width=\columnwidth]{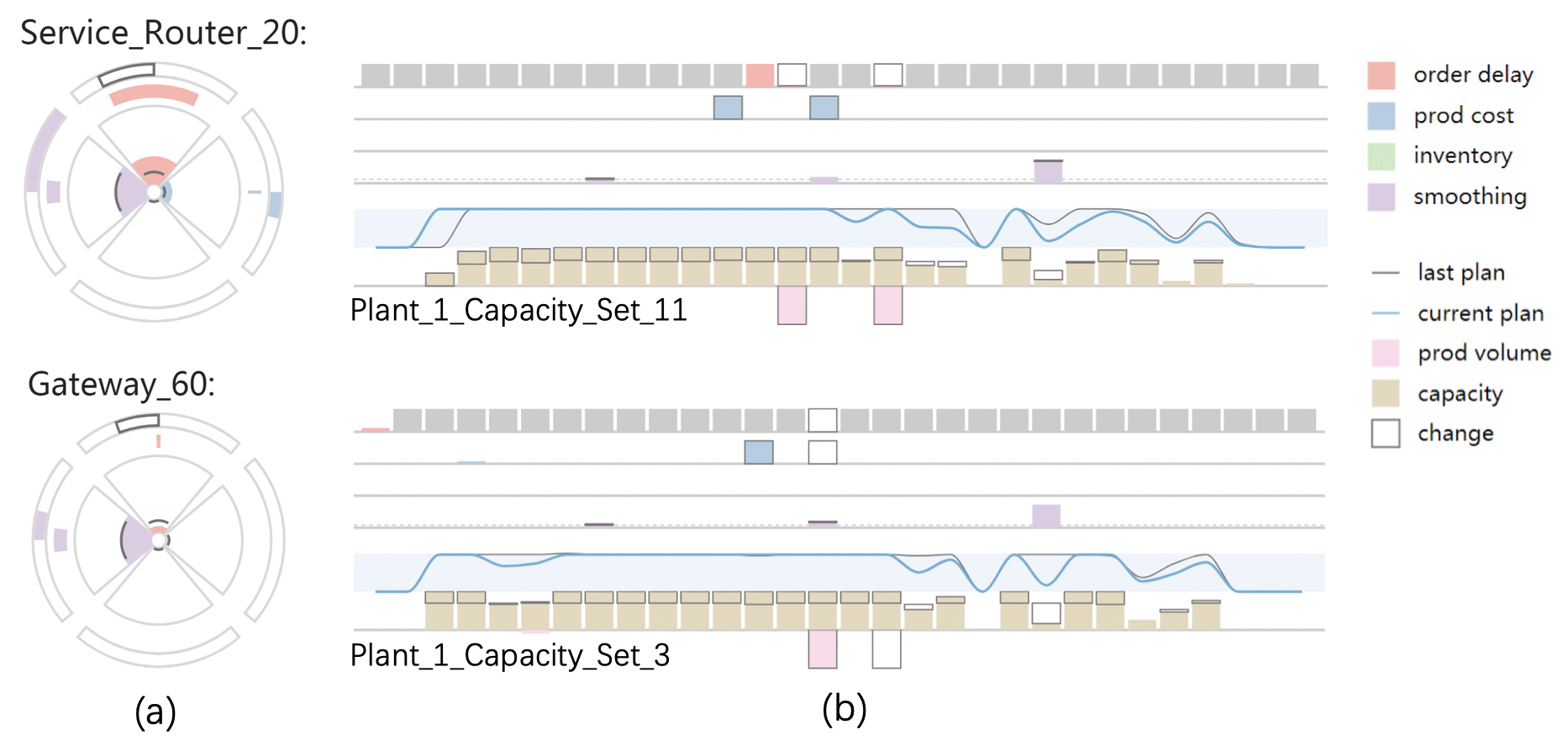}
    \caption{
    The product glyphs (a) and daily production (b) of \textit{Service\_Router\_20} and \textit{Gateway\_60} after increasing the capacity sets by 50\% for 30 days. The order delay rate decreases because the production is increased or finished before the delivery date of the order. 
    }
    \label{fig:case1_after_increasing_capacity_set}
\end{figure}
After experimenting with different increasing amounts, the users decided to increase these two capacity sets by 50\% for 30 days and generated a new plan (\blue{Fig.~\ref{fig:teaser}b$_2$}). To see the difference between the two plans (\textbf{R8}), the users clicked the two glyphs in the overview, and then filtered the parallel coordinates plot to browse the products with decreasing order delay rates. 
After exploring the glyphs of the selected products, they narrowed down to two examples with decreasing order delay rates and increasing smoothing rates, which are shown in \blue{Fig.~\ref{fig:case1_after_increasing_capacity_set}a}. They then clicked the product glyphs to view the production details (\blue{Fig.~\ref{fig:case1_after_increasing_capacity_set}b}). The order delay rate \modified{dropped} because the production \modified{was} increased (\textit{Service\_Router\_20}) and completed before the delivery date (\textit{Gateway\_60}). 

\begin{figure}[tb]
 \centering
 \includegraphics[width=\columnwidth]{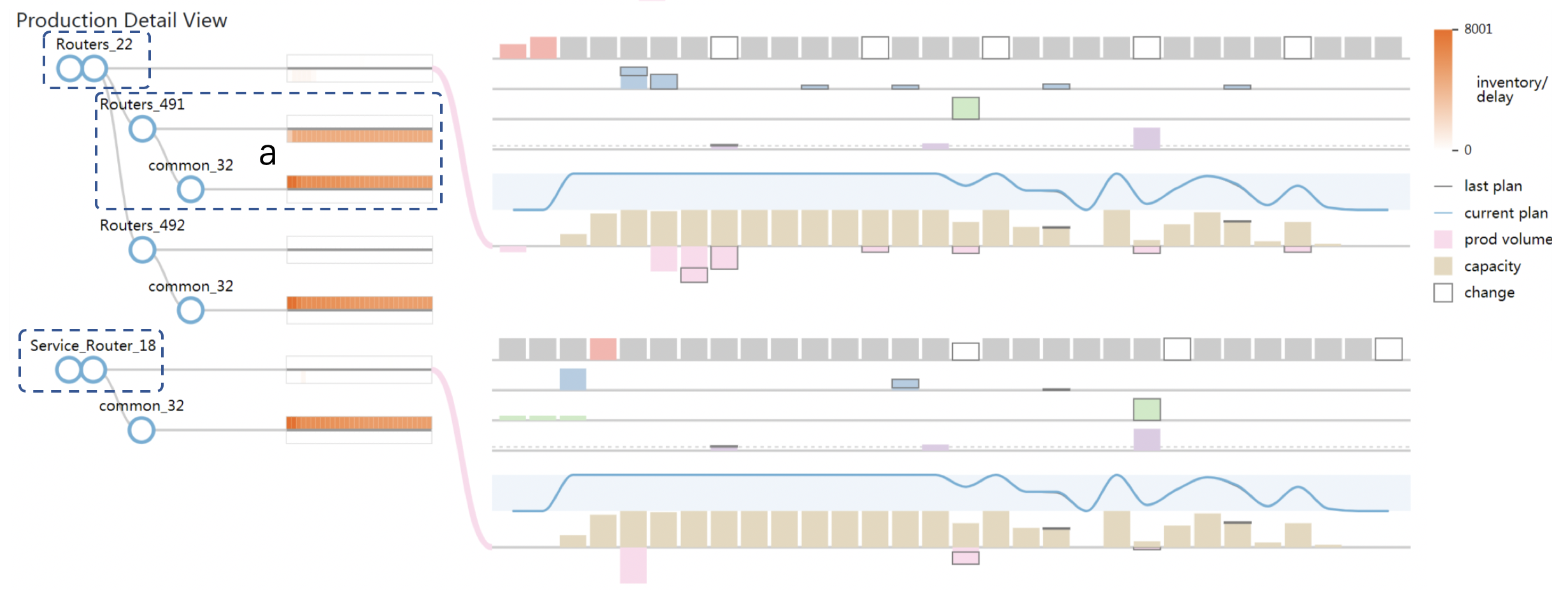}
 \caption{
 The production dependency and daily production of \textit{Routers\_22}, \textit{Routers\_491} and \textit{Service\_Router\_18} after increasing the initial inventory of the raw material \textit{common\_32} from 1000 to 8000. The order delay rate of \textit{Routers\_22} and \textit{Service\_Router\_18} decreases while that of \textit{Routers\_491} is still high. 
 }
 \label{fig:case1_after_increasing_inventory}
\end{figure}

Although the production capacity is sufficient in the improved production plan, there are still some products with high order delay rates (\blue{Fig.~\ref{fig:teaser}c$_1$}). The users brushed these products and the glyphs (\blue{Fig.~\ref{fig:teaser}c$_2$}) showed that most of the products had a low production cost, a low inventory cost, and a decreasing smoothing rate (\textbf{R2}). 
By clicking the product glyph to further explore the production details in \blue{Fig.~\ref{fig:teaser}d} (\textbf{R4}, \textbf{R5}), the users identified three products, namely, \textit{Routers\_22}, \textit{Routers\_491} and \textit{Service\_Router\_18}, whose production relied on the raw material \textit{common\_32}. They found the capacity set was adequate after Sept. 27 (\blue{Fig.~\ref{fig:teaser}d$_4$}), but the orders were delayed due to the lack of \textit{common\_32} (\blue{Fig.~\ref{fig:teaser}d$_2$}). 
To solve the problem, the users tried to increase the initial inventory of \textit{common\_32}. After several trials, they decided to increase the inventory from 1000 to 8000 since this result \modified{led} to the lowest delay rate.
The result is illustrated in \blue{Fig. \ref{fig:case1_after_increasing_inventory}}, where the order delay of \textit{Routers\_22} and \textit{Service\_Router\_18} has been greatly relieved. 

\begin{figure}[tb]
 \centering
 \includegraphics[width=\columnwidth]{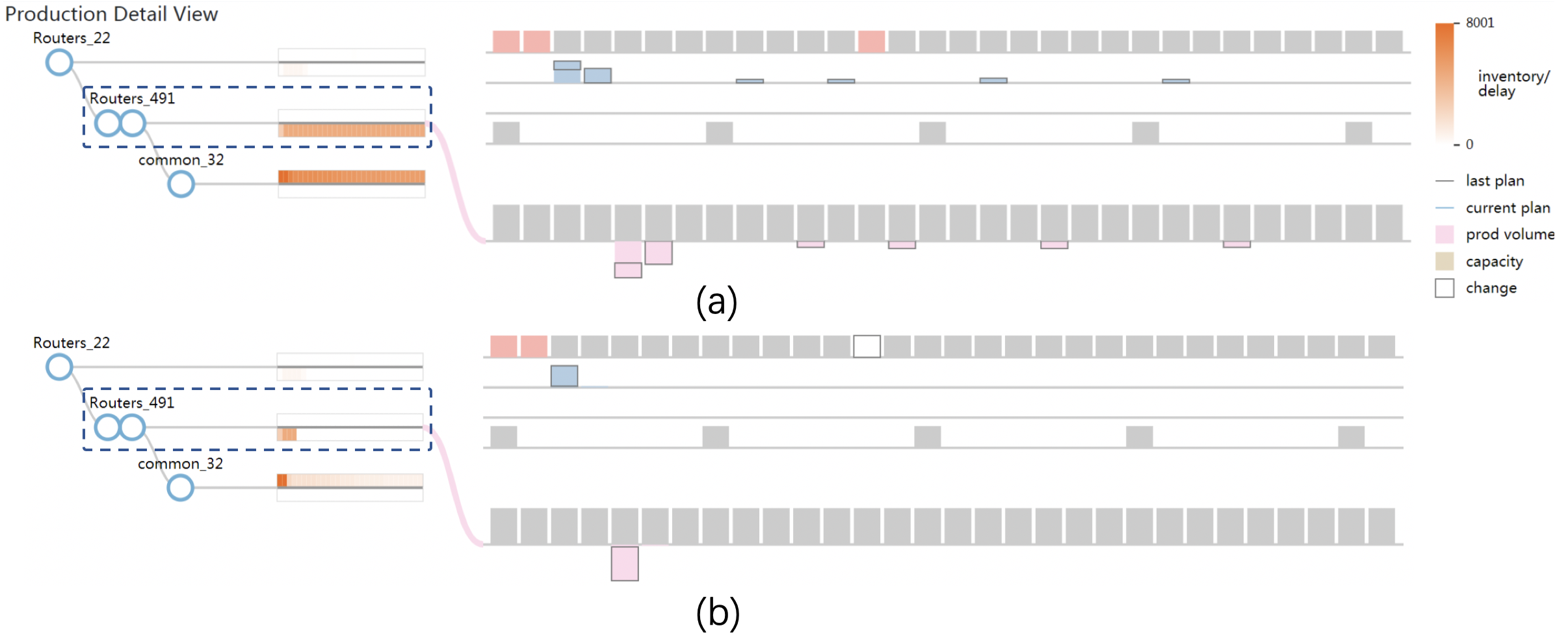}
 \caption{
 (a) The order demand of \textit{Routers\_491} is delayed because of the requirement that \textit{common\_32} can only be used to serve the order of \textit{Routers\_22} and \textit{Service\_Router\_18}. 
 (b) After removing the restriction, the production of \textit{Routers\_491} is increased and the order delay rate drops. 
 }
 \label{fig:case1_after_removing_constraint}
\end{figure}

However, the users noticed that \textit{Routers\_491} still kept a high order delay rate, even though the stock of the raw material \textit{common\_32} was enough (\blue{Fig.~\ref{fig:case1_after_increasing_inventory}a}). Then, the users clicked the heatmap to further browse the daily production of \textit{Routers\_491} (\blue{Fig.~\ref{fig:case1_after_removing_constraint}a}). They observed the production of \textit{Routers\_491} was increased but not used to serve its own order demand. 
One of the users explained that it was caused by a special production constraint called the fixed component requirement, which is generated based on the importance of products and previous experience in coping with raw material shortage. 
In this case, \textit{common\_32} can only be used to produce \textit{Service\_Router\_18} and \textit{Routers\_22}. After they removed the constraint in the planning algorithm, the order delay rate of \textit{Routers\_491} decreased, as shown in \blue{Fig.~\ref{fig:case1_after_removing_constraint}b}. 

In summary, the users showed great interest in {\name} and highly appreciated its capability to gain deep insights into production planning and to fix potential flaws in a production plan.


\subsection{The Quick Response to Unanticipated Incidents in Manufacturing}
\begin{figure}[tb]
 \centering
 \includegraphics[width=\columnwidth]{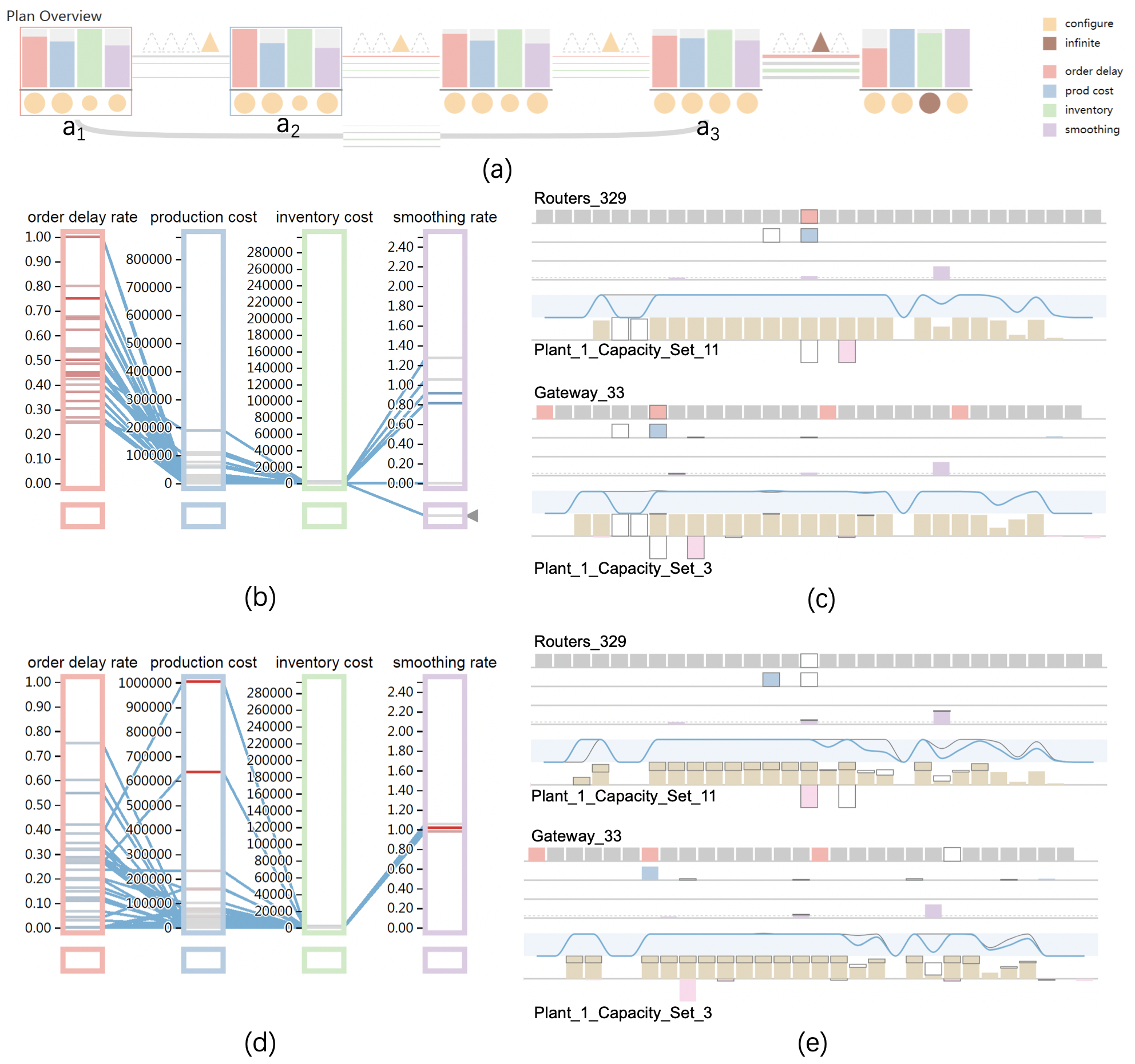}
 \caption{
 (a) The overview of plans generated during the rapid response to typhoon Mangkhut. 
 (b, c) \textit{Plant\_1} is shut down for two days due to the typhoon Mangkhut. The order delay rate of many products is increased since the production is put off. 
 (d, e) After increasing \textit{Plant\_1\_Capacity\_Set\_11} and \textit{Plant\_1\_Capacity\_Set\_3} by 50\% for 30 days, the order delay rate of many products decreases, while the production cost and the smoothing rate of production capacity use increase. It is because the production is increased or advanced. 
 }
 \label{fig:case2}
\end{figure}

\modified{In the second case study, the users were asked to take measures to reduce the adverse influence caused by unanticipated changes in the market or the plant (\textbf{R7}).}
When the users applied {\name} to production planning from Sept. 12 to Oct. 11 in 2018, they found that typhoon Mangkhut \modified{had attacked} the city where \textit{plant\_1} located on Sept. 16 and Sept. 17. \modified{With the advent of the typhoon warning}, they decided to shut down \textit{plant\_1} on the two days. The users simulated this change by adding holidays in the control panel, with the updated plan shown in \blue{Fig.~\ref{fig:case2}a$_2$} (\textbf{R1}). The users then filtered the parallel coordinates plot (\textbf{R3}) and found there were many products with increasing order delay rates (\blue{Fig.~\ref{fig:case2}b}). 
They then brushed the products with increasing order delay rates and decreasing smoothing rates (\textbf{R2}). The control panel shows that the production of the selected products will consume \textit{Plant\_1\_Capacity\_Set\_11} and {Plant\_1\_Capacity\_Set\_3}. The users then clicked the product glyphs and the heatmap in the BOM tree to browse the daily production in related factories (\blue{Fig.~\ref{fig:case2}c}). 
They observed that the production was delayed due to the shortage of the production capacity. 

To handle the adverse influence of the typhoon, the users decided to increase the two capacity sets. They experimented with different increasing values, and finally decided to increase both of the capacity sets by 50\% for 30 days (\blue{Fig.~\ref{fig:case2}a$_3$}), which resulted from the trade-off between the increased production cost and the decreased order delay rate (\blue{Fig.~\ref{fig:case2}d}). 
The users then clicked the product glyphs \modified{again} to view the daily production. As illustrated in \blue{Fig.~\ref{fig:case2}e}, the order delay rate drops since there is enough production capacity to make sure the production can be finished before the delivery date of the order (\textbf{R5}). 
However, there also exist some order demands that cannot be delivered on time, because the delivery date is near the end of the extra holidays, and thus the time left for the production is not enough. 

The users then clicked the initial plan (\blue{Fig.~\ref{fig:case2}a$_1$}) and the plan with increased capacity sets (\blue{Fig.~\ref{fig:case2}a$_3$}) to view their differences (\textbf{R8}). Although the adjustment can relieve the increase of the order delay rate, it comes at the cost of increased smoothing rates. 

Besides the sudden change in manufacturing, our approach also supports the fast response to unanticipated incidents in the market demand and the supply of raw materials, such as urgent orders and returns of raw materials. 

%% file: src/expert_interview.tex
\section{Expert Interviews}
\label{sec_expert_interview}
We also conducted one-on-one interviews with six domain experts from our industry collaborator to evaluate the effectiveness of {\name}.
\modified{
The experts are different from those involved in the design of the system and have never used our system before the interview. Four of them ($E_1$-$E_4$) are experts of production planning algorithm and the other two ($P_1$-$P_2$) are production planning practitioners who are responsible for manually checking the algorithm results and further assigning production tasks to different factories. $P_1$ and $P_2$ are also the target users of the visualization system.
}

The interviews started with a brief introduction to the visual designs, interactions and workflow of our visualization system. 
Also, we showed a usage scenario to the experts to teach them how to operate our system. \modified{The learning process lasted about 25 minutes.}
Then, we asked the experts to freely explore our system by themselves first and further finish a certain task. For example, identify potential problems in a plan and improve it, and re-arrange the production plan in response to a sudden change in the market or production capability.
During this stage, we asked the participants to think aloud
and they could ask questions if they encountered any problems.
Their comments at this stage were also recorded.
After that, we further asked the experts to comment on the design,  the effectiveness and the usage of {\name}.
The whole interviews took around 60 minutes.

Their feedback and suggestions are summarized as follows:

\textbf{Methodology.}
The experts appreciated the method which enables the combination of automatic algorithms and users' domain knowledge for production planning. They reported that the algorithm was often not able to model all the factors for real production planning and they usually needed to manually improve the algorithm results.
$P_1$ commented ``{\name} involves both the algorithm and our experience for production planning, which 
can significantly improve
the production planning efficiency''. The experts also confirmed that the two types of what-if analysis are common and crucial to their daily work. $E_2$ mentioned ``The system makes it easy to identify the adverse influence of the uncertainty in the market and manufacturing''. 
$E_1$ said that {\name} provided guidance for the optimization of production planning and the quick response to unanticipated changes. 

\textbf{Effectiveness.}
The majority of the experts stated that {\name} was beneficial for exploring large-scale production planning data. For instance, 
$P_2$ said that {\name} could greatly reduce his workload as he could quickly check production plans with the help of the plan overview for \modified{a} quick comparison of plans, the product view for fast investigation of product states, and the production detail view for exploring the product dependency and planning details.
In their \modified{current} daily work, they usually need to check details in different tables.
$E_4$ reported that the quick adjustment of production planning in \modified{the} case of unanticipated changes could reduce much loss in real production planning. 
$E_3$ also pointed out that our system could be very helpful for localizing the cause of problems in production plans.
Despite all the positive feedback, the experts and practitioners also mentioned a few limitations and gave us good suggestions on {\name}. 
$E_1$ suggested that the system could further reveal the competition for raw materials and production capacity between products. 
$P_1$ commented that different planners were responsible for different products and it would be better if the system could help filter products for them. 

\textbf{Usability.}
The experts agreed that the visual designs and the interactions in {\name} are well-designed and can facilitate easy exploration of production planning data. 
$P_2$ commented that the recorded optimization history in the plan overview could reduce their mental work and provided an easy way to compare plans. 
After exploring the product view, $E_3$ thought the segmented parallel coordinates plot was efficient in presenting the multiple performance indicators of hundreds of products and the differences between two plans. The experts enjoyed the way of narrowing down the number of \modified{products} for further exploration by filtering, searching and brushing. 
$E_1$ appreciated the design of the skewed tree in the production detail view. He commented ``the alignment of the ending inventory and the delayed order is useful for root reason analysis'' and ``it can be used to illustrate the relationship between factories and assembly lines''. Besides, the experts also stated the production detail view was easy to understand since it adopted visual elements they were familiar with, such as the bar chart, the line chart, and the tree layout. 
Although the visual design is informative, the experts commented that the learning curve of {\name} was a little steep. \modified{However, once they became familiar with the system, they found it powerful in production planning.}

%% file: src/discussion.tex
\section{Discussion}
Our case studies and expert interviews have demonstrated the effectiveness and usability of {\name}. However, there are still several aspects that need further discussion.

\modifiedSecond{\textbf{Generalizability.}}
\modified{We mainly evaluated {\name} on the planning data of factory production. However, it can be extended to other application domains, 
such as the analysis of task assignment of the machines in one plant, vehicle and crew scheduling, and the performance analysis of different scheduling schemes of computer resources (i.e., CPU, memory, threads, and so on).}



\textbf{Algorithm scalability.}
\modified{In} some cases, it may take several seconds for the current production planning algorithm to generate plans, which, however, is still not fast enough and can affect the usability of the system, especially when there are a significantly large number of products and plants.
This issue will be handled in the following perspectives.
First, we can make full use of the computation power of multi-core CPU and GPU to accelerate the production planning algorithm. 
Second, global optimization is utilized each time we invoke the algorithm, even though the change to the configuration data of production planning is only related to a small number of products or plants. Therefore, we expect a local optimization method which can improve the production of related products and in related plants.
Third, we plan to explore other advanced algorithms with better efficiency and further integrate them into {\name}.

\textbf{Visual design scalability.}
Currently, the plan overview only supports the comparison of less than ten plans,
since the production planners usually only need to compare several production plans in their daily practice. 
\modified{
The product view can only support the exploration of hundreds of products.
When there are more products,
filtering strategies can be used here to reduce the number of products to be explored.}
In addition, the production detail view only supports the comparison of several factories, since the domain experts suggest that a product will only be produced in a small number of factories. Furthermore, displaying the detailed production \modified{of} many factories simultaneously \modified{will be} overwhelming.


\textbf{System extensions.}
There are some promising directions to further extend the capability of the {\name} system. 
First, our approach currently does not take the transportation time and cost into consideration. However, the transportation between two plants is also useful for fully  \modified{utilizing} raw materials and child components in the BOM tree. 
Second, as suggested by the domain experts, the visualization system can also be applied to the planning of multiple production lines within a factory. 
Third, the algorithm employed in {\name} can be replaced when more advanced production planning algorithms are available. Also, multiple automatic algorithms of production planning can be further incorporated into {\name}. The system can enable users to interactively compare and select a suitable planning algorithm.

%% file: src/conclusion_and_future_work.tex
\section{Conclusion and Future Work}
In this paper, we propose {\name}, a visual analytics system to support interactive exploration and comparison of production plans in three levels of details. 
Our approach enables the combination of the automatic planning algorithm with the domain expertise of planning practitioners to support two kinds of what-if analyses: the daily optimization of production planning and the quick response to unanticipated changes in the manufacturing process or the market (e.g., a sudden decrease \modified{in} production capacity or raw material supply, and an abrupt increase in market demand).
Thus, our system \modified{is aligned} with the target of implementing smart factories~\cite{kagermann2013recommendations}, including accelerating production planning and quickly adjusting the production plan according to real-time manufacturing data.
We presented two case studies with real-world production planning data and conducted interviews with six domain experts from a world-leading manufacturing company. The results demonstrate the effectiveness and usability of {\name} in exploring, comparing and quickly adjusting production plans. 

In future work, we will consider more performance indicators in production planning, such as the transportation cost and the machine failure rate, and support the fast response to more unanticipated incidents in manufacturing, e.g., traffic jams. 
Also, our approach only shows the summarized production output and capacity use at the factory level. We plan to reveal the production details \modified{of} each production line and machine within a factory.
Furthermore, we would like to integrate more automatic algorithms for production planning into {\name} and it would be interesting to support the comparison of different algorithms and further recommend the most suitable planning algorithm for different production planning tasks.


%% file: src/acknowledgements.tex
\acknowledgments{
We would like to thank the domain experts from Huawei in China for the helpful discussions and the support of user studies. 
Besides, we would like to thank Zezheng Feng for video editing. 
Finally, we would also like to thank the anonymous reviewers for their constructive comments. 
This work is partially supported by grant RGC GRF 16241916 and HK TRS T44-707/16-N.}